\documentclass[aps,prl,twocolumn,superscriptaddress,longbibliography]{revtex4-2}
\usepackage{graphicx}
\usepackage{amssymb,amsmath}
\usepackage{bm}
\usepackage{dcolumn}
\usepackage{float}
\usepackage[OT1]{fontenc} 
\usepackage{url}
\usepackage{mathrsfs}
\usepackage{slashed,comment}
\usepackage{color}
\usepackage{verbatim}
\usepackage{enumitem}
\usepackage{soul,physics}
\usepackage[driverfallback=dvipdfm]{hyperref}
\hypersetup{pdfpagemode=FullScreen,colorlinks=true,breaklinks,urlcolor=blue,linkcolor=blue,citecolor=blue}

\usepackage{subfigure}
\usepackage{amssymb}
\usepackage{bm}
\usepackage{graphicx}
\usepackage{amsmath,amssymb}

\usepackage{pdfpages} 
\usepackage{pgffor} 

\makeatletter
\AtBeginDocument{\let\LS@rot\@undefined}
\makeatother

\def\supplementfilename{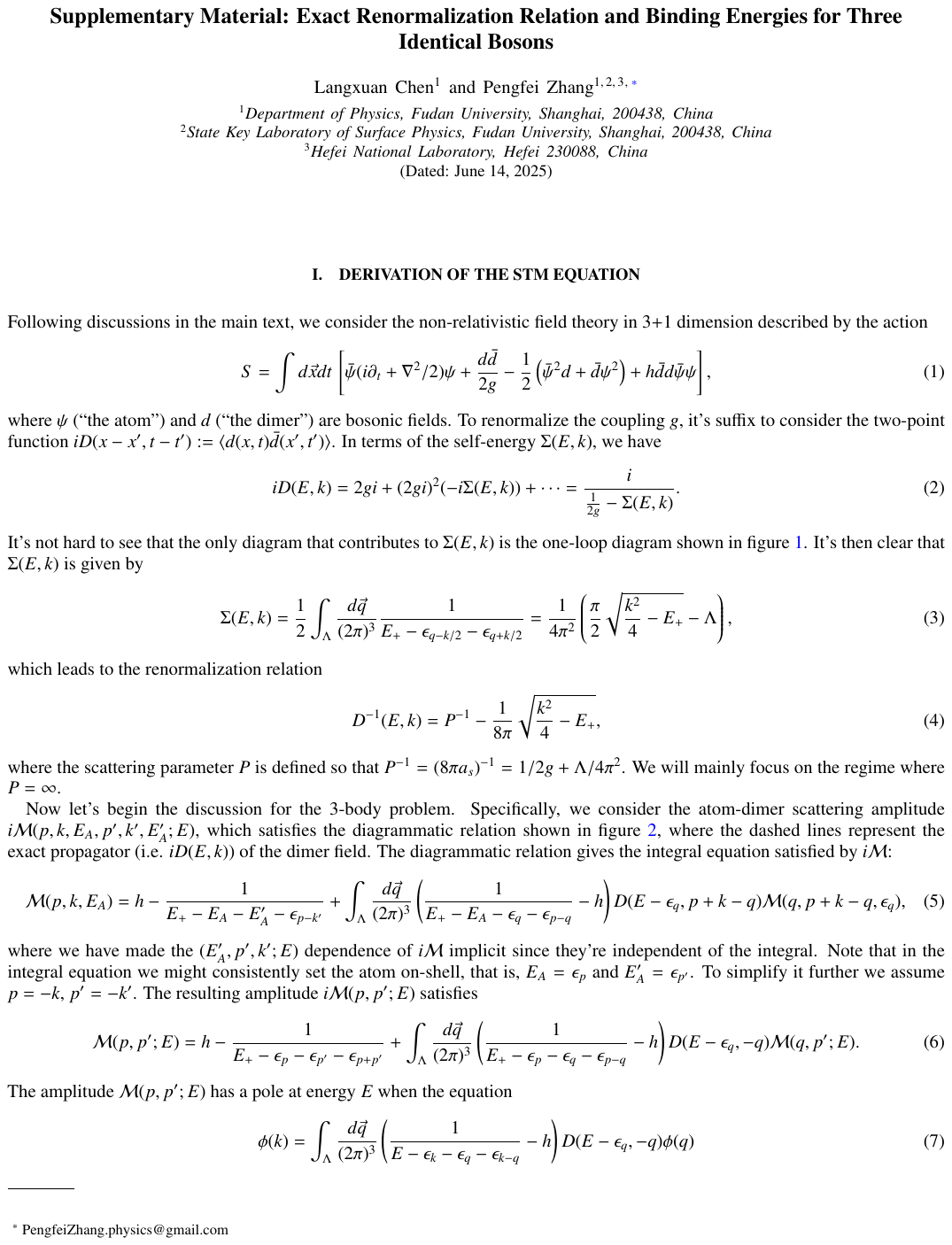}

\pdfximage{\supplementfilename}
\def\numbersupplementpages{\the\pdflastximagepages}

\newif\ifarXiv
\arXivtrue 

\begin{document}
 
  \title{Exact Renormalization Relation and Binding Energies for Three Identical Bosons}

  \author{Langxuan Chen}
  \affiliation{Department of Physics, Fudan University, Shanghai, 200438, China}

  \author{Pengfei Zhang}
  \thanks{PengfeiZhang.physics@gmail.com}
  \affiliation{Department of Physics, Fudan University, Shanghai, 200438, China}
  \affiliation{State Key Laboratory of Surface Physics, Fudan University, Shanghai, 200438, China}
  \affiliation{Hefei National Laboratory, Hefei 230088, China}

  \date{\today}

  \begin{abstract}
  In the low-energy limit, non-relativistic particles with short-range interactions exhibit universal behavior that is largely independent of microscopic details. This universality is typically described by effective field theory, in which the two-body interaction is renormalized to a single parameter-the scattering length. For systems of identical bosons, the three-body problem reveals the Efimov effect, a novel phenomenon proposed that necessitates the introduction of an additional three-body parameter. However, the exact relation between this three-body parameter, the coupling constants in the effective field theory, and the binding energies of Efimov states remains unresolved. In this Letter, we address this question through a comprehensive analysis of the Skorniakov-Ter-Martirosian equation with a finite cutoff. We establish an exact renormalization relation for the three-body parameter and determine its connection to the energies of Efimov bound states. These results are validated through high-precision numerical simulations. We expect our findings to be of fundamental interest across various fields, including atomic, nuclear, condensed matter, and particle physics, and to have broad applications in both few-body and many-body physics.
  \end{abstract}
    
  \maketitle

  \emph{ \color{blue}Introduction.--} 
  Universality refers to phenomena in which systems with diverse microscopic details exhibit identical behavior in the low-energy limit. A prominent example is the scattering of non-relativistic identical bosons with short-range interactions, which has wide applications in atomic, nuclear, condensed matter, and particle physics \cite{Hammer:2005bp,Bedaque:1998km,Nielsen:2001hbm,Ferlaino2010FortyYO,Braaten:2006vd,Hammer:2010kp,Nishida:2012hf,Naidon:2016dpf,Kievsky:2021ghz}. When the typical momentum of the bosons is much smaller than the inverse interaction range, the universal behaviors can be captured by an effective field theory with the Lagrangian density \cite{Hammer:2005bp,Bedaque:1998km}: 
  \begin{equation}
  \begin{aligned}
    \mathcal{L}=&\bar\psi\Big(i\partial_t+\frac{\nabla^2}{2}\Big)\psi+\frac{d\bar{d}}{2g}-\frac{\bar{\psi}^2d+\bar{d}\psi^2}{2}+h\bar d\bar \psi\psi d+...
    \end{aligned}
  \end{equation}
  Here, we set the boson mass $m = 1$ for conciseness. $d$ is a dimer field introduced for convenience, commonly referred to as a dibaryon in nuclear systems. The ellipsis denotes additional terms involving more derivatives and/or fields. The effective field theory is valid only in the low-energy limit, which requires the introduction of a momentum cutoff $\Lambda$. To make experimentally relevant predictions, it is necessary to establish renormalization relations that connect the cutoff-dependent coupling constants to physical parameters.

  The renormalization of $g$ is well established by analyzing the two-body problem \cite{Hammer:2005bp}. We consider two incoming bosons with total energy $E$ and total momentum $\bm{k}$, both of which are conserved during the scattering process. The two-body $T$-matrix can be obtained by evaluating the dressed Green's function of the dimer field, $D(E, \bm{k})$, which is given by
  \begin{equation}
    D^{-1}(E,\bm{k})=\frac{1}{2g}+\frac{\Lambda}{4\pi^2}-\frac{1}{8\pi}\sqrt{\frac{k^2}{4}-E_+}.
  \end{equation}
  Here, we introduce $E_+=E+i0^+$. The cutoff dependence of $g$ should be chosen to cancel the explicit cutoff dependence in the theory. This leads to the following renormalization relation ${(4\pi a_s)}^{-1}=g^{-1}+{\Lambda}/{2\pi^2}$, where $a_s$ is the scattering length \cite{landau2013quantum}, the sole parameter that characterizes the universal properties of dilute many-body systems from two-body physics \cite{zhai2021ultracold}. At $a_s = \infty$, the two-body sector exhibits emergent scale invariance, and the system is said to be on resonance.

  Next, we turn to the three-body problem, in which the boson-dimer coupling $h$ plays a role. We focus on systems at two-body resonance $a_s=\infty$ \footnote{Note that the renormalization relation derived in this work also applies to large but finite values of $a_s$.}. For three-body bound states with total energy $E<0$ and momentum $\bm{k} = \bm{0}$, diagrammatic calculations show that the $s$-wave bound-state wavefunction $\phi(k)$ satisfies the Skorniakov-Ter-Martirosian (STM) equation \cite{Skorniakov:1957kgi,Hammer:2005bp}:
  \begin{equation}
    \int_0^\Lambda \frac{q^2dq}{\sqrt{q^2-4E/3}}\Big(K(k,q,E)+h\Big)\phi(q)=\lambda\phi(k),
    \label{eq:Efimov3dE}
\end{equation}
where we have introduced $\lambda=\sqrt3 \pi/8$ and $$K(k,q,E)=\frac{1}{2kq}\ln\left(\frac{k^2+q^2+kq-E}{k^2+q^2-kq-E}\right).$$
The values of $E$ for which Eq.~\eqref{eq:Efimov3dE} admits non-trivial solutions correspond to the energies of three-body bound states. The result reveals a remarkable phenomenon known as the Efimov effect, discovered by Vitaly Efimov in 1970 \cite{EFIMOV1970563} and experimentally observed in both cold atom systems \cite{2006Natur.440..315K,PhysRevLett.112.190401,PhysRevLett.112.250404,PhysRevLett.113.240402} and the helium trimer \cite{doi:10.1126/science.aaa5601}. The presence of the Efimov effect originates from the breaking of continuous scale invariance down to discrete scale invariance due to the UV cutoff. As a consequence, Eq.~\eqref{eq:Efimov3dE} admits an infinite tower of three-body bound states, characterized by the scaling relation $E_n / E_{n+1} = \exp(2\pi/s_0)$ as $n \rightarrow \infty$. Here, $s_0\approx 1.00624$ is determined by solving $$\mathcal{G}(s_0)\equiv \frac{\pi \sinh(\pi s_0/6)}{s_0 \cosh(\pi s_0/2)} =\lambda.$$ The discrete scale invariance is reflected in the universal log-periodicity of $\phi(k)$ in the intermiediate regime $\Lambda \gg k\gg \sqrt{|E|}$ \cite{Hammer:2005bp}:
\begin{equation}\label{eq:waveuniversal}
\phi(k)\propto k^{-1}\cos\left({s_0}\ln (k/\Lambda_*)\right).
\end{equation}
We will present a derivation of this result later. Here, $\Lambda_*$ determines the phase shift of the three-body wavefunction and is referred to as the three-body parameter. 

There are two important questions concerning the three-body parameter $\Lambda_*$.
(1) What is the exact renormalization relation among the coupling constant $h$, the three-body parameter $\Lambda_*$, and the cutoff $\Lambda$?
(2) Given the three-body parameter $\Lambda_*$, what are the precise binding energies of the Efimov bound states $E_n$ in the low-energy limit? Although these questions play a fundamental role in understanding the univserality of the identical bosons for both few-body and many-body systems, their exact answers remain unknown owing to the complexity of Eq.~\eqref{eq:Efimov3dE} at finite $\Lambda$ or non-zero $E$. Regarding the first question, in the pioneering work \cite{Bedaque:1998kg}, an analytical approximation for the renormalization relation was derived under the assumption that Eq.~\eqref{eq:waveuniversal} holds for $k \lesssim \Lambda$. By dimensional analysis, one introduces $h = H/\Lambda^2$. Imposing cutoff-independence of the low-energy wavefunction then gives
\begin{equation}\label{eqn:literature1}
H\approx \alpha \left(\frac{1+s_0 \tan{\varphi}}{1-s_0 \tan{\varphi}}\right),
\end{equation}
where we define $\varphi = s_0 \ln(\Lambda_*/\Lambda)$, and the analytical approximation yields $\alpha = 1$. However, numerical simulations with an accuracy of about $10^{-3}$ later suggested a more precise value of $\alpha \approx 0.879$ \cite{PhysRevLett.106.153005}. For the second question, we introduce the binding momentum $\kappa_*$ via $s_0\ln \kappa_*\equiv s_0\ln \sqrt{|E_n|}\ \ \text{mod}\ \pi$ for $n\gg 1$. Previous numerical simulations yield \cite{Hammer:2005bp}
\begin{equation}
s_0 \ln(\Lambda_*/\kappa_*)\approx 0.971\ \text{mod}\ \ \pi.
\end{equation}

  \emph{ \color{blue}Summary of the Results.--} In this letter, we present exact analytical results that address both questions, derived from a detailed analysis of Eq.~\eqref{eq:Efimov3dE} at finite $\Lambda$ or non-zero $E$. Our findings are summarized as follows:

  \vspace{5pt}

  (1). The exact renormalization relation reads
  \begin{equation}
    H=\alpha\left(\frac{1+s_0 \tan{(\varphi-\varphi_0)}}{1-s_0 \tan{(\varphi-\varphi_0)}}\right),
    \label{eq:result11}
  \end{equation}
  with $\alpha$ and $\varphi_0$ determined by integrals
  \begin{equation}
  \begin{aligned}
\alpha&=2\lambda(1+s_0^2)\exp\!\left(\frac{1}{\pi}\int_{-\infty}^\infty d\omega\,\frac{\ln F(\omega)}{1+\omega^2}\right),\\
\varphi_0&=\arg\left[\frac{1}{1+is_0}\exp\!\left(\frac{s_0}{\pi i}\,\text{P}.\!\int_{0}^\infty d\omega\,\frac{\ln F(\omega)}{\omega^2-s_0^2}\right)\right].
\end{aligned}
\label{eq:result12}
\end{equation}
Here, P$.$ denotes the principle value of the integral and 
\begin{equation}
    F(\omega)=\frac{1}{s_0^2-\omega^2}\left(\lambda^{-1}\mathcal{G}(\omega)-1\right).
\end{equation}
By evaluating the integrals numerically, we find that
\begin{equation}
\alpha\approx 0.87866\ \ \ \ \ \ \varphi_0\approx 0.05281.
\end{equation}

(2). The binding energies of Efimov states are
\begin{equation}\label{eq:result2}
s_0 \ln(\Lambda_*/\kappa_*)=-s_0\ln \left(\sqrt{3}e^{-\pi/2s_0}\right)\ \ \text{mod}\ \pi.
\end{equation}
Direct evaluation yields $-s_0\ln \left(\sqrt{3}e^{-\pi/2s_0}\right) \approx 1.01807$. 

\vspace{5pt}

In addition, our analysis predicts the exact form of the bound state wavefunction for $n \gg 1$, which holds for arbitrary $k \in (0, \Lambda)$ and is consistent with Eq.~\eqref{eq:waveuniversal} in the regime $\Lambda \gg k \gg \sqrt{|E|}$. For example, when $k\ll \Lambda$, we have $\phi(k)\propto k^{-1}\sin\!\left(s_0\,\text{arcsinh}\!\left(k/\sqrt{-4E/3}\right)\right).$ Therefore, our results provide a complete characterization of the low-energy behavior of three identical bosons at the two-body resonance. In the remainder of this manuscript, we first present numerical demonstrations of our results and compare them with approximations reported in the literature. We then outline the main steps of our proof, with full technical details provided in the supplementary material \cite{SM}. Finally, we conclude with a discussion of potential applications to universal properties of few-body and many-body systems from three-body physics.

  \begin{figure}[t]
    \centering
    \includegraphics[width=0.8\linewidth]{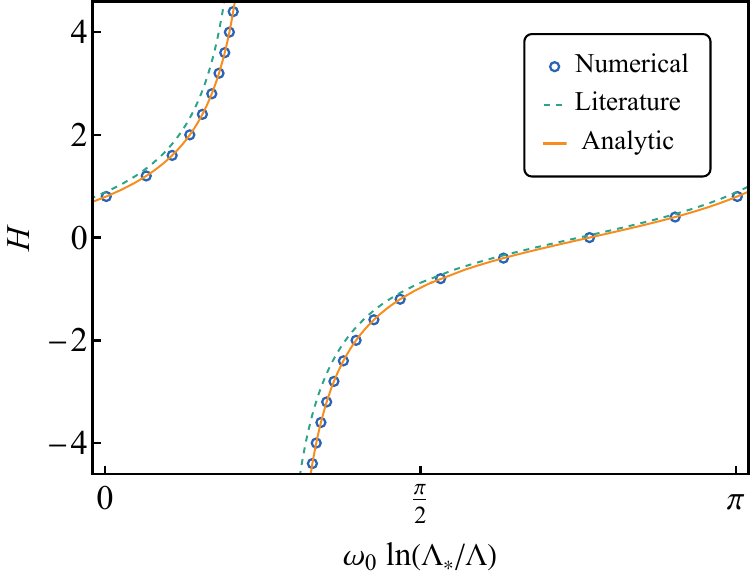}
    \caption{ Numerical validation of the exact renormalization relation. Dark blue dots represent numerical results obtained by discretizing the integral equation~\eqref{eq:Efimov3dE}. The dashed green line shows the previously reported result in the literature (Eq.\eqref{eqn:literature1} with $\alpha = 0.879$), while the solid orange line corresponds to our exact renormalization relation in Eq.\eqref{eq:result11} and \eqref{eq:result12}. The results show excellent agreement between our analytical prediction and the numerical data. }
    \label{fig1}
  \end{figure}

\emph{ \color{blue}Numerical Validation.--} To validate our results in Eqs.~\eqref{eq:result11}, \eqref{eq:result12}, and \eqref{eq:result2}, we numerically solve the STM equation by discretizing the function $\phi(k)$ and express \eqref{eq:Efimov3dE} as a matrix equation. To improve numerical accuracy, we employ a uniform discretization of the variable $\tau = \text{arcsinh}(k/\sqrt{-4E/3})$. This choice corresponds to a uniform discretization in $k$ for $k \ll \sqrt{-E}$ and in $\ln k$ for $k \gg \sqrt{-E}$, thereby capturing the log-periodic behavior predicted by Eq.\eqref{eq:waveuniversal}. We further apply Simpson’s 1/3 rule to evaluate the integral within each sub-interval. With a discretization of $10^3$ points, this method achieves a numerical accuracy of approximately $10^{-6}$. The bound-state energy and corresponding wavefunction are then determined by searching for the value of $E$ at which the discretized equation admits a non-trivial solution. Finally, we determine the value of $\kappa_*$ using its definition\eqref{eq:waveuniversal}. 
 
  \begin{figure}[t]
    \centering
    \includegraphics[width=0.8\linewidth]{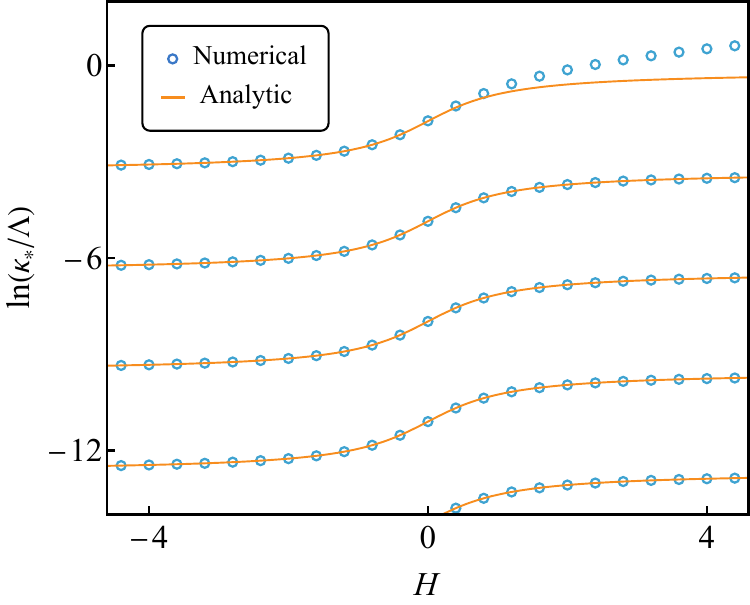}
    \caption{ Numerical validation of the binding energies of Efimov bound states in the limit $n \rightarrow \infty$. Dark blue dots represent numerical results obtained by discretizing the integral equation~\eqref{eq:Efimov3dE}, while the solid red line shows the analytical prediction based on Eqs.~\eqref{eq:result11} and \eqref{eq:result2}. The comparison reveals good agreement between our analytical prediction and the numerical data starting from the second-lowest bound state. }
    \label{fig2}
  \end{figure}

Our numerical results for the renormalization relation between $H$ and $\ln(\Lambda_*/\Lambda)$ are presented in Fig. \ref{fig1}, alongside comparisons with both the previously reported result in the literature (Eq.~\eqref{eqn:literature1} with $\alpha = 0.879$) and our theoretical predictions in Eqs.~\eqref{eq:result11} and \eqref{eq:result12}. The numerical data exhibit excellent agreement with our analytical results, with discrepancies below $10^{-7}$. Next, we present numerical results for the binding energies $|E_n|$ as a function of $H$ in Fig.~\ref{fig2}. The theoretical prediction, based on Eq.~\eqref{eq:result11} and \eqref{eq:result2}, is expected to hold in the low-energy limit $n \rightarrow \infty$. The results show good agreement beginning with the second-lowest bound state, where the relative error is $|\Delta E_1 / E_1| \lesssim 10^{-3}$. For the third-lowest bound state, the error further decreases to $|\Delta E_2 / E_2| \lesssim 10^{-6}$. These results clearly demonstrate the validity of the theoretical prediction for both the exact renormalization relation and the binding energies of the Efimov bound states.

\begin{figure}[t]
    \centering
    \includegraphics[width=0.8\linewidth]{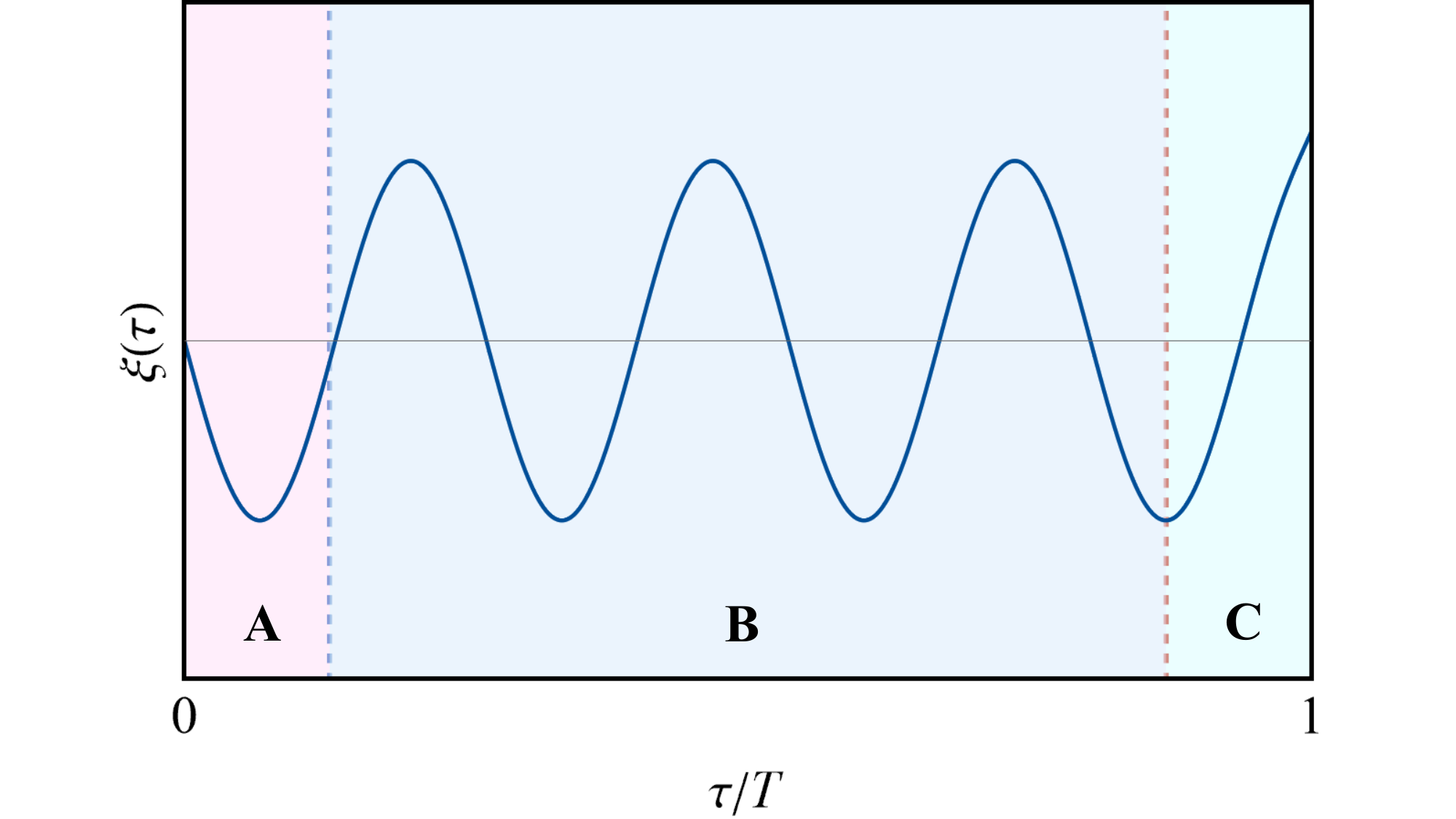}
    \caption{An illustration for Regime I and Regime II. Here, we use the variable $\tau = \text{arcsinh}(k/\sqrt{-4E/3})$ and $T=\text{arcsinh}(\Lambda/\sqrt{-4E/3})$. The full domain is divided into three regions: Region A ($k \lesssim \sqrt{-E}$), Region B ($\sqrt{-E}\ll k \ll \lambda$), and Region C ($k\lesssim \Lambda$). Regime I consists of Regions B and C, while Regime II includes Regions A and B. The blue line represents a typical solution $\xi(\tau)$. }
    \label{illustrate}
  \end{figure}

  \emph{ \color{blue}Sketch of the Proof.--} We first define two different regimes for the analysis of the STM equation \eqref{eq:Efimov3dE}. For shallow bound states, the problem exhibits a separation of energy scales. The renormalization relation is determined by the high-momentum behavior of the wavefunction, which can be analyzed in \textbf{Regime I}, where $k \gg \sqrt{|E|}$. In this regime, we may set $E = 0$ and impose the boundary condition given in Eq.~\eqref{eq:waveuniversal} at $k \ll \Lambda$. On the other hand, the binding energy is governed by the low-momentum behavior of the wavefunction, corresponding to \textbf{Regime II}, where $k \ll \Lambda$. In this regime, we can take the limit $\Lambda \rightarrow \infty$ and match the boundary condition \eqref{eq:waveuniversal} at $k \gg \sqrt{|E|}$. We now proceed with a detailed analysis of both regimes. An illustration for different regimes is presented in Fig.~\ref{illustrate}
 
  \textbf{Regime I}.-- After setting $E = 0$ and introducing the new parameter $t \equiv \ln \frac{\Lambda}{k} \in (0, \infty)$ and the wavefunction $\phi(k) = k^{-1} \xi(t)$, we arrive at the equation:
  \begin{equation}
    \int_{0}^\infty ds\,\left(G(t-s)+He^{-(t+s)}\right)\xi(s)=\lambda\xi(t),
    \label{eq:simplifiedmain}
  \end{equation}
  Here, we have introduced 
  \begin{equation}\label{eqn:Gdef}
  G(t)=\frac{1}{2}\ln\left(\frac{\cosh(t)+1/2}{\cosh(t)-1/2}\right).
  \end{equation}
  The appearance of $G(t - s)$ in Eq.~\eqref{eq:simplifiedmain} reflects the scale invariance at the two-body resonance. It is straightforward to check that $\mathcal{G}(\omega)=\int_{-\infty}^\infty dt\,e^{i\omega t}G(t)$ is the Fourier transform of $G(t)$. To warm up, we consider the low-energy solution in the regime $t \gg 1$ (or equivalently $k \ll \Lambda$). Since $G(t)$ decays exponentially at large $|t|$, we can safely extend the domain of integration to $-\infty$, and the equation becomes translationally invariant. The solution takes the form $\xi(t) \propto \cos(s_0 t + \varphi)$, where $s_0$ is determined by the condition $\mathcal{G}(s_0) = \lambda$. By comparing this solution with Eq.~\eqref{eq:waveuniversal}, we identify the phase shift $\varphi = s_0 \ln(\Lambda_*/\Lambda)$, as introduced in previous sections. This provides an explanation for the emergence of log-periodic behavior of the wavefunction within our framework. 

  Determining the renormalization relation requires solving Eq.\eqref{eq:simplifiedmain} for arbitrary $t$. An important observation is that for $H = 0$, Eq.\eqref{eq:simplifiedmain} reduces to a Wiener-Hopf integral equation, a well-studied class of equations in the mathematical literature \cite{noble1959methods}. Our approach to solving Eq.~\eqref{eq:simplifiedmain} generalizes the method developed for solving the Wiener-Hopf equation, with analyticity playing a central role. As a first step, we extend the domain of $\xi(t)$ to the entire real axis by defining $\xi(t) \equiv 0$ for $t < 0$. However, Eq. \eqref{eq:simplifiedmain} is no longer valid for $t<0$. Therefore, it is necessary to introduce an error term $e(t)$, so that $e(t)\equiv0$ for $t> 0$ and $\xi(t)$ satisfies the extended equation
  \begin{equation}
  \begin{aligned}
    \int_{-\infty}^\infty ds\Big(G(t-s)+H \theta(t)&e^{-(t+s)}\Big)\xi(s)\\&=\lambda \left(\xi(t)+e(t)\right).
    \end{aligned}
  \end{equation}
  This equation takes a simple form in the frequency domain, which reads
  \begin{equation}
    (\lambda^{-1}\mathcal{G}(\omega)-1)\tilde\xi(\omega)+\frac{H\tilde \xi(i)}{\lambda(1-i\omega)}=\tilde e(\omega),
    \label{eq:fouriermaineq}
  \end{equation}
  where $\tilde \xi(\omega)$ and $\tilde e(\omega)$ are Fourier transform of $\xi(t)$ and $e(t)$. Since $e(t) = 0$ for $t > 0$ and decays exponentially for $t<0$ and $|t| \gg 1$, its Fourier transform $\tilde e(\omega)$ is analytic throughout the lower-half complex plane, including the real axis. In contrast, the wavefunction $\tilde\xi(\omega)$ is analytical on the upper-half plane. However, it possesses poles on the real axis, which give rise to the periodic oscillations observed at large $t$.

  To utilize these analytical properties, we first examine the structure of the factor $(\lambda^{-1}\mathcal{G}(\omega)-1)$. This function decays monotonically along the real axis as $|\omega|$ increases and vanishes at $\pm s_0$. Consequently, the auxiliary function $F(\omega)=(\lambda^{-1}\mathcal{G}(\omega)-1)/(s_0^2-\omega^2)$ is strictly positive on the real axis, and is analytic and zero-free within a narrow strip $B:=\{\omega\in\mathbb C\,|-a\leq\Im(\omega)\leq a\}$ for any positive constant $a<2$. This allows us to make the splitting \cite{SM}
  \begin{equation}
  F(\omega)=F_+(\omega)F_-(\omega),
  \end{equation}
  where $F_+(\omega)$ and $F_-(\omega)$ are analytic and zero-free in the upper- and lower-half complex planes, respectively. In particular, their definitions within the strip $B$ can be written explicitly using the contour integral:
  \begin{equation}
    F_{\pm}(\omega)\equiv\exp\left(\pm\frac{1}{2\pi i}\int_{\text{Im}~z=\mp a}dz\,\frac{\ln F(z)}{z-\omega}\right).
    \label{eq:Fpm}
  \end{equation}
  
  Next, let us define
  \begin{equation}
    f(\omega)\equiv(s_0^2-\omega^2)(1-i\omega)F_+(\omega)\tilde \xi(\omega).
    \label{f1}
  \end{equation}
  It is straightforward to verify that, by definition, $f(\omega)$ is analytic on the upper-half plane, including the real axis. Furthermore, the equation \eqref{eq:fouriermaineq} becomes
  \begin{equation}
  f(\omega)F_-(\omega)=(1-i\omega)\tilde e(\omega)-\frac{Hf(i)}{2\lambda F_+(i)(1+s_0^2)}.
     \label{f2}
  \end{equation}
  Therefore, the function $f(\omega)$ is also analytic in the lower half-plane. Consequently, $f(\omega)$ is analytic over the entire complex plane. In the supplementary material \cite{SM}, we further show that $\frac{|f(\omega)|}{1+|\omega|}$ is bounded. Thus, by applying Liouville's theorem, we conclude that $f(\omega) = \kappa - i \omega$ is a linear function of $\omega$ \footnote{Note that an overall factor is unimportant since the STM equation is homogeneous.}. 

  We can determine $\kappa$ setting $\omega=-i$ in Eq. \eqref{f2}. This leads to 
  \begin{equation}
  \kappa=\frac{\alpha-H}{\alpha+H},\ \ \ \ \alpha=2\lambda(1+s_0^2)|F_+(i)|^2.
  \end{equation}
  By applying the contour-integral representation \eqref{eq:Fpm} and smoothly deforming the integration contour to the real axis, this definition of $\alpha$ matches Eq. \eqref{eq:result12} presented in summary section. This relation, together with the definition \eqref{f1}, completely determines the wavefunction $\tilde{\xi}(\omega)$ and enables the derivation of the renormalization relation: We compute the residue $A$ of $\tilde{\xi}(\omega)$ at $\omega = -s_0$, which leads to the asymptotic expansion $\xi(t) \sim A e^{i s_0 t} + \text{c.c.}$. Therefore, we can identify $\arg A = \varphi$. On the other hand, straightforward calculation gives 
  \begin{equation}
  A\propto(s_0-i\kappa)\beta,\ \ \ \ \beta=\frac{\pi}{s_0(1+is_0)F_+(-s_0)}.
  \end{equation}
  Introducing $\text{arg}~\beta=\varphi_0$ leads to $\tan(\varphi-\varphi_0)=-\kappa/s_0$. Finally, expressing $\kappa$ in terms of $H$ leads to the exact renormalization \eqref{eq:result11}, and the contour deformation yields the expression \eqref{eq:result12} for $\varphi_0$.

  \textbf{Regime II}.-- After taking the limit $\Lambda \rightarrow \infty$ and changing variables to $\tau = \text{arcsinh}(k/\sqrt{-4E/3})$, the STM equation \eqref{eq:Efimov3dE} becomes:
  \begin{equation}
  \begin{aligned}
    \lambda\xi(\tau)&=\int_0^{\infty}d\sigma ~K(\tau,\sigma)\xi(\sigma),\\
    K(\tau,\sigma)&=G(\tau-\sigma)-G(\tau+\sigma),
    \end{aligned}
    \label{eq:Efimov3dwithEsimp}
\end{equation}
where $G$ is given by Eq. \eqref{eqn:Gdef}. Setting $\tau = 0$ and using the fact that $G(\tau)$ is an even function of $\tau$, we find that this equation is consistent with extending the domain of $\xi(\tau)$ by defining $\xi(-\tau) = -\xi(\tau)$. Consequently, the equation \eqref{eq:Efimov3dwithEsimp} is equivalent to
\begin{equation}
\lambda\xi(\tau)=\int_{-\infty}^{\infty}d\sigma ~G(\tau-\sigma)\xi(\sigma).
\end{equation}
Therefore, the solution is the equation is exactly given by $\xi(\tau)\propto \sin(s_0\tau)$. To match the boundary condition at $\tau \gg 1$, we make the approximation that $\tau\approx \ln (k/\sqrt{|E|/3})$. This leads to 
\begin{equation}
\xi(k)\propto \sin\Big(s_0 \ln (k/\sqrt{|E|/3})\Big). \ \ \ \ (k\gg \sqrt{|E|})
\end{equation}
Finally, we match this result to the boundary condition \eqref{eq:waveuniversal} set by the three-body parameter, which gives
\begin{equation}
s_0 \ln (k/\sqrt{|E|/3})-\frac{\pi}{2}={s_0}\ln (k/\Lambda_*)\ \ \text{mod}\ \pi.
\end{equation}
By identifying $s_0 \ln \kappa_* = s_0 \ln \sqrt{|E|} \ \ \text{mod} \ \pi$, we arrive at the result given by Eq. \eqref{eq:result2}.

  \emph{ \color{blue}Discussions.--} 
In this letter, we establish the exact renormalization relation and binding energies for the three-body problem of identical bosons. To derive the renormalization relation, our main tool is a generalization of the Wiener-Hopf method, which exploits the analyticity of the integral kernel. The exact binding energies are obtained by leveraging the special structure of the integral equation after a suitable change of variables. Given the fundamental relevance of the effective field theory of identical bosons across various areas of physics, we expect our results to have broad applicability. 

We highlight a series of intriguing directions. Although our focus is on non-relativistic identical bosons, the analysis can be directly extended to derive exact renormalization relations in other scenarios where the Efimov effect emerges—such as highly imbalanced Fermi gases \cite{Hammer:2005bp}, systems in mixed dimensions \cite{Nishida:2011ew,PhysRevLett.101.170401,PhysRevA.79.060701}, or systems with long-range interactions \cite{Sun:2025gew}. More importantly, our results pave the way for a precise characterization of both few-body and many-body phenomena rooted in three-body physics, with applications including three-body correlations in Bose polarons \cite{PhysRevLett.119.013401}, the virial expansion of atomic gases \cite{PhysRevA.92.062716}, and universal relations \cite{PhysRevLett.112.110402,PhysRevLett.106.153005,PhysRevA.86.053633}. In all these contexts, the renormalization relation plays a crucial role in understanding universal behaviors.

\vspace{5pt}
\textit{Acknowledgement.} 
We thank Shuyan Zhou and Shuo Zhang for helpful discussions. This project is supported by the Innovation Program for Quantum Science and Technology 2024ZD0300101, the Shanghai Rising-Star Program under grant number 24QA2700300, and the NSFC under grant 12374477.

\bibliography{Efimov_Renormalization.bbl}

\ifarXiv


\section*{Supplementary material}

# Supplementary Material: Exact Renormalization Relation and Binding Energies for Three Identical Bosons

Langxuan Chen[1] and Pengfei Zhang[1, 2, 3, *]

[1]*Department of Physics, Fudan University, Shanghai, 200438, China*
[2]*State Key Laboratory of Surface Physics, Fudan University, Shanghai, 200438, China*
[3]*Hefei National Laboratory, Hefei 230088, China*
(Dated: June 14, 2025)

## I. DERIVATION OF THE STM EQUATION

Following discussions in the main text, we consider the non-relativistic field theory in 3+1 dimension described by the action

$$S = \int d\vec{x} dt \left[ \bar{\psi}(i\partial_t + \nabla^2/2)\psi + \frac{d\bar{d}}{2g} - \frac{1}{2} \left( \bar{\psi}^2 d + \bar{d}\psi^2 \right) + h\bar{d}d\bar{\psi}\psi \right], \tag{1}$$

where $\psi$ ("the atom") and $d$ ("the dimer") are bosonic fields. To renormalize the coupling $g$, it's suffix to consider the two-point function $iD(x - x', t - t') := \langle d(x,t)\bar{d}(x',t') \rangle$. In terms of the self-energy $\Sigma(E,k)$, we have

$$iD(E,k) = 2gi + (2gi)^2(-i\Sigma(E,k)) + \cdots = \frac{i}{\frac{1}{2g} - \Sigma(E,k)}. \tag{2}$$

It's not hard to see that the only diagram that contributes to $\Sigma(E,k)$ is the one-loop diagram shown in figure 1. It's then clear that $\Sigma(E,k)$ is given by

$$\Sigma(E,k) = \frac{1}{2} \int_\Lambda \frac{d\vec{q}}{(2\pi)^3} \frac{1}{E_+ - \epsilon_{q-k/2} - \epsilon_{q+k/2}} = \frac{1}{4\pi^2} \left( \frac{\pi}{2} \sqrt{\frac{k^2}{4} - E_+} - \Lambda \right), \tag{3}$$

which leads to the renormalization relation

$$D^{-1}(E,k) = P^{-1} - \frac{1}{8\pi} \sqrt{\frac{k^2}{4} - E_+}, \tag{4}$$

where the scattering parameter $P$ is defined so that $P^{-1} = (8\pi a_s)^{-1} = 1/2g + \Lambda/4\pi^2$. We will mainly focus on the regime where $P = \infty$.

Now let's begin the discussion for the 3-body problem. Specifically, we consider the atom-dimer scattering amplitude $i\mathcal{M}(p, k, E_A, p', k', E_A'; E)$, which satisfies the diagrammatic relation shown in figure 2, where the dashed lines represent the exact propagator (i.e. $iD(E,k)$) of the dimer field. The diagrammatic relation gives the integral equation satisfied by $i\mathcal{M}$:

$$\mathcal{M}(p, k, E_A) = h - \frac{1}{E_+ - E_A - E_A' - \epsilon_{p-k'}} + \int_\Lambda \frac{d\vec{q}}{(2\pi)^3} \left( \frac{1}{E_+ - E_A - \epsilon_q - \epsilon_{p-q}} - h \right) D(E - \epsilon_q, p + k - q) \mathcal{M}(q, p + k - q, \epsilon_q), \tag{5}$$

where we have made the $(E_A', p', k'; E)$ dependence of $i\mathcal{M}$ implicit since they're independent of the integral. Note that in the integral equation we might consistently set the atom on-shell, that is, $E_A = \epsilon_p$ and $E_A' = \epsilon_{p'}$. To simplify it further we assume $p = -k$, $p' = -k'$. The resulting amplitude $i\mathcal{M}(p, p'; E)$ satisfies

$$\mathcal{M}(p, p'; E) = h - \frac{1}{E_+ - \epsilon_p - \epsilon_{p'} - \epsilon_{p+p'}} + \int_\Lambda \frac{d\vec{q}}{(2\pi)^3} \left( \frac{1}{E_+ - \epsilon_p - \epsilon_q - \epsilon_{p-q}} - h \right) D(E - \epsilon_q, -q) \mathcal{M}(q, p'; E). \tag{6}$$

The amplitude $\mathcal{M}(p, p'; E)$ has a pole at energy $E$ when the equation

$$\phi(k) = \int_\Lambda \frac{d\vec{q}}{(2\pi)^3} \left( \frac{1}{E - \epsilon_k - \epsilon_q - \epsilon_{k-q}} - h \right) D(E - \epsilon_q, -q)\phi(q) \tag{7}$$

---

* PengfeiZhang.physics@gmail.com

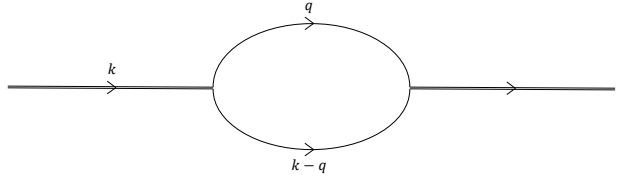

FIG. 1. One Loop Contribution to $d\bar{d}$ Propagator

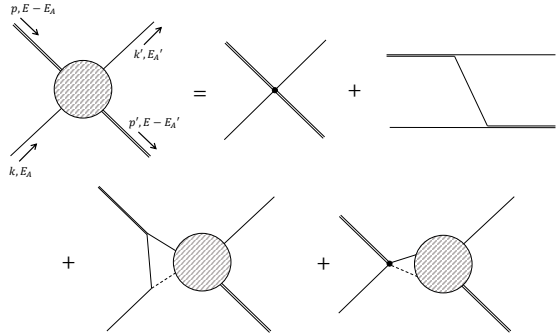

FIG. 2. Diagrammatic relation satisfied by atom-dimer scattering amplitude

has a non-trivial solution $\phi(k)$. When this happens, we have 3-body bound state with energy $E$. Note that we have removed the $i\epsilon$ dependence bound states have energy $E < 0$ and there is no need for regularization. Focusing on $s$-wave bound states, we have

$$\phi(k) = -\frac{1}{2\pi^2}\int_0^\Lambda dq\, q^2 \left(P^{-1} - \frac{1}{8\pi}\sqrt{3q^2/4 - E}\right)^{-1}\left(\frac{1}{2kq}\ln\left(\frac{k^2 + q^2 + kq - E}{k^2 + q^2 - kq - E}\right) + h\right)\phi(q), \quad k, q > 0. \tag{8}$$

If we assume a infinite scattering parameter $P = \infty$, then

$$\phi(k) = \frac{8}{\sqrt{3}\pi}\int_0^\Lambda dq\, \frac{q^2}{\sqrt{q^2 - 4E/3}}\left(\frac{1}{2kq}\ln\left(\frac{k^2 + q^2 + kq - E}{k^2 + q^2 - kq - E}\right) + h\right)\phi(q), \tag{9}$$

which is the STM equation.

## II. THE ZERO ENERGY SOLUTION

In the STM equation (9), when $k \gg \sqrt{|E|}$, we may effectively set $E = 0$. The resulting equation reads

$$\phi(k) = \frac{8}{\sqrt{3}\pi}\int_0^\Lambda dq\, q\left(\frac{1}{2kq}\ln\left(\frac{k^2 + q^2 + kq}{k^2 + q^2 - kq}\right) + h\right)\phi(q), \tag{10}$$

which is scale invariant. Setting $\phi(k) = k^{-1}\xi(k)$ and choosing a new parameter $t \equiv \ln(\Lambda/k)$, we arrived at the equation

$$\int_0^\infty ds\, \left(G(t-s) + He^{-(t+s)}\right)\xi(s) = \lambda\xi(t), \quad t > 0 \tag{11}$$

where $\lambda = \sqrt{3}\pi/8$, $H(\Lambda) = \Lambda^2 h(\Lambda)$ and

$$G(t) := \frac{1}{2}\ln\left(\frac{\cosh(t) + 1/2}{\cosh(t) - 1/2}\right) \tag{12}$$

is the kernel of the integral equation. Choosing an appropriate integration contour, it not hard to show that

$$\mathcal{G}(\omega) \equiv \int_{-\infty}^{\infty} dt\, G(t) e^{i\omega t} = \frac{\pi \sinh(\pi\omega/6)}{\omega \cosh(\pi\omega/2)}. \tag{13}$$

To discuss the solution for (10), note that for $H = 0$, we have

$$\int_0^{\infty} ds\, G(t - s)\xi(s) = \lambda\xi(t), \quad t > 0, \tag{14}$$

which is the famous Wiener-Hopf integral equation. Our method for solving equation (11) is basically a slight variation of the Wiener-Hopf method. Although the kernel $G(t)$ is already known in our problem, we shall solve (11) for generic even kernel $G(t)$ that decays exponentially for large $t$. In addition, we shall also assume that $\mathcal{G}(\omega)$ is positive and decreasing for $\omega > 0$, so that the equation $\mathcal{G}(s_0) = \lambda$ has a unique positive solution $s_0 > 0$. We wish to deal with the general kernel because equation (11) could also describe other bound state problem (e.g. the higher partial wave bound states in our model), for which $G(t)$ might be different.

### A. Qualitative Features

Before we proceed to solving (11), let's first make some comments on the qualitative properties of the solution. Note that physically we're mainly interested in the parameter regime where $k \ll \Lambda$, that is, $e^t \gg 1$. Since the kernel $G(t)$ decays exponentially for large $|t|$, we see that for large $t$, $\xi(t)$ satisfies the equation

$$\int_{-\infty}^{\infty} ds\, G(t - s)\xi(s) = \lambda\xi(t). \tag{15}$$

As a result, for large $t$, a generic real solution might be approximated as $\xi(t) \sim A\cos(s_0 t + \varphi)$, where $s_0 > 0$ is determined by the condition $\int_{-\infty}^{\infty} ds\, G(t) e^{is_0 t} = \mathcal{G}(s_0) = \lambda$. Assuming that $\mathcal{G}(\omega)$ is positive and decreasing for $\omega > 0$ with $\mathcal{G}(0) > \lambda$, the equation has a unique positive solution $s_0 > 0$. For the specific kernel (12) we have $s_0 \approx 1.00624$. The "phase shift" $\varphi$ is determined by the equation in the range $t \sim O(1)$. If we switch back to use the original variable $k$, the above asymptotic expansion implies that for $k \ll \Lambda$,

$$\phi(k) \approx k^{-1}\cos(s_0 \ln(k/\Lambda_*)), \tag{16}$$

where the 3-body parameter $\Lambda_*$ is defined by the relation $s_0 \ln(\Lambda_*/\Lambda) = \varphi \mod \pi$. If we're only interested in the low energy behavior of the wave function $\phi(k)$, then $\varphi$ is the only important physical parameter. Therefore, to renormalize $H(\Lambda)$ we shall determine the relation between $H(\Lambda)$ and $\varphi$. When $\Lambda$ changes, $H(\Lambda)$ should be tuned so that the low energy parameter (i.e. $\varphi$ or $\Lambda_*$) remains unchanged.

### B. Solution for Rational Kernel

In this section we begin to explore the solutions. Following the general method for solving Wiener-Hopf equations, we extend equation (11) to the whole real line $\mathbb{R}$, and defining $\xi(t) \equiv 0$ for $t < 0$. It's clear that for $t < 0$, equation (11) is no longer valid. It is therefore necessary to introduce an error term $e(t)$, so that $e(t) \equiv 0$ for $t > 0$ and

$$\int_0^{\infty} ds\, G(t - s)\xi(s) = \lambda e(t), \quad t < 0. \tag{17}$$

Combining the analysis above, we have the equation

$$\int_{\mathbb{R}} ds\left(G(t - s) + H\theta(t)e^{-(t+s)}\right)\xi(s) = \lambda\left(\xi(t) + e(t)\right), \quad t \in \mathbb{R}. \tag{18}$$

Fourier Transform the above equation, we obtain

$$(\lambda^{-1}\mathcal{G}(\omega) - 1)\tilde{\xi}(\omega) + \frac{H\tilde{\xi}(i)}{\lambda(1 - i\omega)} = \tilde{e}(\omega), \tag{19}$$

where $\mathcal{G}(\omega) = \int dt\, e^{i\omega t} G(t)$ is the Fourier transform of $G(t)$, and similarly for $\tilde{\xi}(\omega)$ and $\tilde{e}(\omega)$. Since we assume that $G(t)$ decays exponentially for large $|t|$, we know that $e(t)$ must decays exponentially for large $|t|$. As a result, the Fourier-Transformed result $\tilde{e}(\omega)$ is analytic in the lower-half plane (In fact, it is analytic for $\text{Im}(\omega) < \alpha$ with some $\alpha > 0$). It is essentially the analyticity of $\tilde{e}(\omega)$ that determines the solution on the L.H.S. For simplicity, we first deal with a special class of kernels $G(t)$ (i.e. "rational kernel") for which the problem is easy. Specifically, a kernel $G(t)$ is defined to be *rational* if

$$G(t) = \sum_i c_i e^{-ip_i|t|}, \quad \text{Im}(p_i) < 0. \tag{20}$$

The Fourier transformed kernel is

$$\mathcal{G}(\omega) = -\sum_i \frac{2ic_iq_i}{q_i^2 - \omega^2}. \tag{21}$$

Since $G(t)$ is an even function, $\mathcal{G}(\omega)$ is even as well. Since we've assumed that there exists a unique $s_0$ so that $\mathcal{G}(s_0) = \lambda$, we have

$$\lambda^{-1}\mathcal{G}(\omega) - 1 = (s_0^2 - \omega^2)\frac{\prod_i(\omega - s_i)}{\prod_j(\omega - p_j)}, \tag{22}$$

where $p_i$ are the poles of $\mathcal{G}(\omega)$, and $s_i$ are the zeros of the function $\lambda^{-1}\mathcal{G}(\omega) - 1$ with non-zero imaginary parts.

### *1. The Solution*

Now let's solve (19). For simplicity, we first deal with the case $H = 0$. The equation is

$$\frac{(s_0^2 - \omega^2)\prod_i(\omega - s_i)}{\prod_j(\omega - p_j)}\tilde{\xi}(\omega) = \tilde{e}(\omega). \tag{23}$$

Since $\tilde{e}(\omega)$ is analytic on the lower half-plane, the poles $p_j$ with $\text{Im}(p_j) < 0$ must be canceled by the zeros of the function $\tilde{\xi}(\omega)$. What's more, $\tilde{\xi}(\omega)$ could only have poles at $\pm s_0 - i\epsilon$ or $s_i$ with $\text{Im}(s_i) < 0$. A natural conjecture for the correct solution is given by

$$\tilde{\xi}(\omega) = \left(\frac{i}{\omega_+^2 - s_0^2}\right)\frac{\prod_{\text{Im}(p_j)<0}(\omega - p_j)}{\prod_{\text{Im}(s_i)<0}(\omega - s_i)}, \tag{24}$$

and it's not hard to check that it is indeed the correct solution. Note that for large $|\omega|$ we have $|\tilde{\xi}(\omega)| \sim O(1/|\omega|)$, which makes Jordan's Lemma applicable. To make sure that the corresponding time-domain solution $\xi(t)$ is real, we shall show that $\tilde{\xi}(\omega)^* = \tilde{\xi}(-\omega^*)$. Note that

$$\prod_{\text{Im}(p_j)<0}(\omega - p_j)^* = \prod_{\text{Im}(p_j)<0}(\omega^* - p_j^*) = \prod_{\text{Im}(p_j)<0}(\omega^* + p_j). \tag{25}$$

The last equality follows from the fact that $\mathcal{G}(\omega)^* = \mathcal{G}(-\omega^*)$, which means $p \mapsto -p^*$ is a permutation of the poles $p$ that preserves $\text{Im}(p)$. For the same reason we have

$$\prod_{\text{Im}(s_i)<0}(\omega - s_i)^* = \prod_{\text{Im}(s_i)<0}(\omega^* + s_i). \tag{26}$$

Since $\lambda^{-1}\mathcal{G}(\omega) - 1 \to -1$ for $|\omega| \to \infty$, we have $\#\{p_j | \text{Im}(p_j) < 0\} = \#\{s_i | \text{Im}(s_i) < 0\} + 1$. Combine the above we obtain the desired result $\tilde{\xi}(\omega)^* = \tilde{\xi}(-\omega^*)$. If we assume the boundness of $\xi(t)$, it is possible to prove mathematically that (24) is the only solution of the given integral equation by showing that other choices of $\tilde{\xi}(\omega)$ and $\tilde{e}(\omega)$ all exhibits undesired behavior as $|\omega| \to \infty$. A rigorous proof for the uniqueness of the solution will be given when we discuss generic kernel $G(t)$ in the next section.

Now we turn to the case $H \neq 0$. The equation is

$$\frac{(s_0^2 - \omega^2) \prod_i (\omega - s_i)}{\prod_j (\omega - p_j)} \tilde{\xi}(\omega) + \frac{H\tilde{\xi}(i)}{\lambda(1 - i\omega)} = \tilde{e}(\omega). \tag{27}$$

The new term is associated with a pole $\omega = -i$ on the lower half-plane, which must being canceled by contributions from $\tilde{\xi}(\omega)$. Motivated by the solutions in the case $H = 0$, we're led to the new solution

$$\tilde{\xi}(\omega) = \frac{i(\kappa - i\omega) \prod_{\mathrm{Im}(p_j)<0}(\omega - p_j)}{(\omega_+^2 - s_0^2)(1 - i\omega) \prod_{\mathrm{Im}(s_i)<0}(\omega - s_i)}, \tag{28}$$

where we have introduced the factor $\frac{1}{1-i\omega}$, and $\kappa$ is a parameter to be determined. Requiring the cancellation of the pole at $\omega = -i$, we have

$$\frac{i(\kappa - 1) \prod_{\mathrm{Im}(s_i)<0}(i - s_i)}{\prod_{\mathrm{Im}(p_j)<0}(i - p_j)} - \frac{iH(\kappa + 1) \prod_{\mathrm{Im}(p_j)<0}(i - p_j)}{2\lambda(1 + s_0^2) \prod_{\mathrm{Im}(s_i)<0}(i - s_i)} = 0. \tag{29}$$

By some elementary algebra, the above equation is equivalent to $\alpha(\kappa - 1) + H(\kappa + 1) = 0$, where

$$\alpha = -\frac{2\lambda(1 + s_0^2) \prod_{\mathrm{Im}(s_i)<0}(i - s_i)^2}{\prod_{\mathrm{Im}(p_j)<0}(i - p_j)^2} > 0. \tag{30}$$

The solution for $\kappa$ is

$$\kappa = \frac{\alpha - H}{\alpha + H}. \tag{31}$$

Since $\kappa$ is real, it's not hard to check the condition $\tilde{\xi}(\omega)^* = \tilde{\xi}(-\omega^*)$ for (28).

### 2. The Phase Shift

In our previous analysis, we have concluded that the only low-energy observable of the solution $\xi(t)$ is the phase-shift $\varphi$. Having established the full-solution $\tilde{\xi}(\omega)$, it's now easy to extract $\varphi$. For large $t$, we have the asymptotic expansion $\xi(t) \sim Ae^{is_0 t} + c.c.$ As a result, we know that $\varphi = \arg(A) + k\pi$. Applying Jordan's Lemma to the Fourier inversion formula $\xi(t) = \frac{1}{2\pi} \int d\omega \, e^{-i\omega t} \tilde{\xi}(\omega)$, we obtain $A = (s_0 - i\kappa)\beta$, where

$$\beta = \frac{i \prod_{\mathrm{Im}(p_j)<0}(s_0 + p_j)}{2s_0(1 + is_0) \prod_{\mathrm{Im}(s_i)<0}(s_0 + s_i)}. \tag{32}$$

Defining $\varphi_0 := \arg \beta$, we have $s_0 \tan(\varphi - \varphi_0) = -\kappa$. Combine the results above we have

$$H = \alpha \left( \frac{1 + s_0 \tan(\varphi - \varphi_0)}{1 - s_0 \tan(\varphi - \varphi_0)} \right), \tag{33}$$

which explicitly established the relation between $H$ and $\varphi$. Although the above relation is established for rational kernels only, we expect it to be true for generic kernel $G(t)$ since $\mathcal{G}(\omega)$ might always be approximated by a rational function. Therefore, the above analysis already establishes the renormalization relation between $H$ and $\varphi$ claimed in the main text, except that we have to determine the parameters $\alpha$ and $\varphi_0$ by fitting. To determine $\alpha$ and $\varphi_0$ analytically, we need the solution for a generic kernel, which is discussed in the next section.

### C. Solutions for Generic Kernel

The above analysis for rational kernels $G(t)$ has the virtue of being explicit. However, in most practical applications, the rational assumption (20) is at best an approximation. In addition, the solution $\tilde{\xi}(\omega)$ is increasingly complicated as we increasing the number of terms in (20), which is undesirable if we wish to solve a generic kernel $G(t)$ by rational approximations of $\mathcal{G}(\omega)$. In this section we shall therefore focusing on generic kernels. As before, we assume that the kernel $G(t)$ is even, positive, $C^\infty$ on $\mathbb{R}_+$ and decays exponentially for large $|t|$. (Note that we only assume that $G(t)$ is $C^\infty$ on $\mathbb{R}_+$, and it could be non-differentiable at $t = 0$) We also assume that $\mathcal{G}(\omega)$ is positive and monotonically decreasing for $\omega > 0$, with a unique solution $s_0 > 0$ such that $\lambda^{-1}\mathcal{G}(s_0) - 1 = 0$.

### 1. Structure of Analyticity

Like what we did in the rational case, we shall first analyze the properties of the function $\lambda^{-1}\mathcal{G}(\omega) - 1$. Define $F(\omega) := \frac{\lambda^{-1}\mathcal{G}(\omega)-1}{s_0^2-\omega^2}$, we have $F(\omega) > 0$ for $\omega \in \mathbb{R}$. The properties of the kernel $G(t)$ guarantees the existence of a positive constant $a > 0$, such that $F(\omega)$ is analytic and *zero-free* on the strip $B := \{\omega \in \mathbb{C} \mid -a \le \mathrm{Im}(\omega) \le a\}$. Since $F > 0$ for $\omega \in \mathbb{R}$, $\ln F(\omega)$ is uniquely defined on the strip $B$ with $\ln F(\omega) \in \mathbb{R}$ for $\omega \in \mathbb{R}$.

### 2. Solution

Before we make any further discussion, let's first revisit our solution for a rational kernel. In the solution, the functions

$$F_\pm(\omega) := \pm i \frac{\prod_{\pm \mathrm{Im}(s_i)<0}(\omega - s_i)}{\prod_{\pm \mathrm{Im}(p_j)<0}(\omega - p_j)} \tag{34}$$

are of central importance. They are related to each other by the relations $F_+(-\omega) = F_-(\omega)$ and $F_+(\omega)^* = F_-(\omega^*)$. In particular, we have

$$F_+(\omega)F_-(\omega) = \frac{\prod_i(\omega - s_i)}{\prod_j(\omega - p_j)} = F(\omega) \tag{35}$$

for a rational kernel $G$. The mathematical meaning of $F_\pm(\omega)$ are clear: they decomposes $F(\omega)$ into the product of two factors that are analytic and zero-free in the upper and lower half-plane respectively. Let's assume that the splitting is also achievable for a generic kernel $G(t)$, that is, we have With the help of the decomposition, equation (19) reads

$$(s_0^2 - \omega^2)F_+(\omega)F_-(\omega)\tilde{\xi}(\omega) + \frac{H\tilde{\xi}(i)}{\lambda(1-i\omega)} = \tilde{e}(\omega). \tag{36}$$

Defining

$$f := (s_0^2 - \omega^2)(1 - i\omega)F_+\tilde{\xi}, \tag{37}$$

so that

$$(1 - i\omega)\tilde{e} = \left(fF_- + \frac{Hf(i)}{2\lambda F_+(i)(1+s_0^2)}\right). \tag{38}$$

By the analyticity of $\tilde{e}(\omega)$ and $\tilde{\xi}(\omega)$, it's clear that $f$ is holomorphic on the full complex plane. In addition, since $\tilde{e}$ has no pole on the lower half-plane, $(1-i\omega)\tilde{e} = 0$ for $\omega = -i$. It leads to the equality

$$2\lambda(1 + s_0^2)F_+(i)F_-(-i)f(-i) + Hf(i) = 0. \tag{39}$$

To proceed further we need to know more about the properties of the functions $F_\pm(\omega)$. Here we present an explicit construction. For generic $F(\omega)$, we define

$$F_\pm(\omega) := \exp\left(\pm\frac{1}{2\pi i}\int_{\mathrm{Im}(z)=\mp a}dz\,\frac{\ln F(z)}{z-\omega}\right), \tag{40}$$

where we assume $-a < \mathrm{Im}(\omega) < a$. One may easily check that the conditions $F_+F_- = F$, $F_+(-\omega) = F_-(\omega)$ and $F_+(\omega)^* = F_-(\omega^*)$ are satisfied by the construction (40). In addition, $F_+(\omega)$ might be extended to function that is holomorphic and zero-free for $\mathrm{Im}(\omega) > -a$, and the same is true for $F_-(\omega)$ with $\mathrm{Im}(\omega) < a$. As it was mentioned in the main text, we shall prove that $f(\omega)$ is linear in $\omega$. The basic idea is to bound $f(\omega)$ on the upper and lower half-plane respectively. On the upper-half plane, we have $|F_+(\omega)| \sim O(1/|\omega|)$ and $|\tilde{\xi}(\omega)| \sim O(1/|\omega|)$ (The proof are given below). Therefore, combine with (37) we have $|f(\omega)| \sim O(|\omega|)$ on the upper-half plane. Similarly, from (38) we might show that $|f(\omega)| \sim O(|\omega|)$ on the lower-half plane as well. Therefore, $f(\omega) \sim O(|\omega|)$ on the entire complex plane. Since $f$ is holomorphic, it must be a linear function of $\omega$. Now we present the rigorous proof. First, we need the result of a special integral:

$$\pm\frac{1}{2\pi i}\int_{\mathrm{Im}(z)=\mp a}dz\left(\frac{\ln(z^2+b^2)}{z-\omega}\right) = \ln(\omega \pm ib) \mp \frac{i\pi}{2}, \quad b > a. \tag{41}$$

Since $F(\omega) = \frac{\lambda^{-1}\mathcal{G}(\omega)-1}{s_0^2-\omega^2}$, we know that $\ln\left[(\omega^2+b^2)F(\omega)\right] \to 0$ for $\mathrm{Re}(\omega) \to \pm\infty$ on the strip $B$ since $\mathcal{G}(\omega) \to 0$ in that case. As a result, for $\mathrm{Im}(\omega) > -a$, the integral

$$\frac{1}{2\pi i}\int_{\mathrm{Im}(z)=-a} dz\left(\frac{\ln F(z)}{z-\omega}\right) \tag{42}$$

is dominated by $-\frac{1}{2\pi i}\int_{\mathrm{Im}(z)=-a} dz\,\frac{\ln(\omega^2+b^2)}{z-\omega}$, plus corrections that are uniformly bounded by some constant $C > 0$, and similar is true for the integral on $\Gamma^-$. In another words, we have

$$\pm\frac{1}{2\pi i}\int_{\mathrm{Im}(z)=\mp a} dz\left(\frac{\ln F(z)}{z-\omega}\right) = -\ln(\omega \pm ib) + h_\pm(\omega), \tag{43}$$

where $|h_+(\omega)| < C$ for $\mathrm{Im}(\omega) > -a$ and $|h_-(\omega)| < C$ for $\mathrm{Im}(\omega) < a$, respectively (Note that different choices of $b > a$ gives different $h_\pm$).

To constrain the behavior of $f$, we also need to analyze the behavior of $\tilde{\xi}(\omega)$ and $\tilde{e}(\omega)$ on the complex plane. Assuming that $\xi(t)$ and $\xi'(t)$ are bounded for $t > 0$. Since the kernel $G(t)$ decays exponentially at large time separation, we know that $e(t)$ and $e'(t)$ decay exponentially for $t \to -\infty$. As a result, we may find $\epsilon > 0$, so that $|e(t)e^{-2\epsilon t}|$ and $|e'(t)e^{-2\epsilon t}|$ are bounded function for $t < 0$. For $\mathrm{Im}\,\omega \geq \epsilon$, we have

$$\tilde{\xi}(\omega) = -\frac{1}{i\omega}\left(\xi(0) + \int_0^\infty dt\,\xi'(t)e^{i\omega t}\right). \tag{44}$$

Since $|\omega^{-1}| \leq \frac{1+\epsilon^{-1}}{1+|\omega|}$ for $\mathrm{Im}\,\omega \geq \epsilon$, we have

$$|\tilde{\xi}(\omega)| \leq \frac{C_1}{1+|\omega|}, \quad \mathrm{Im}(\omega) \geq \epsilon \tag{45}$$

where we define

$$C_1 := (1+\epsilon^{-1})\left(|\xi(0)| + \epsilon^{-1}\sup_{t>0}|\xi'(t)|\right) < \infty. \tag{46}$$

Similarly, for $\tilde{e}(\omega)$, assuming that $\mathrm{Im}\,\omega \leq \epsilon$, Then

$$\tilde{e}(\omega) = \int_{-\infty}^0 dt\,e^{i(\omega-2i\epsilon)t}(e(t)e^{-2\epsilon t}) = \frac{1}{i(\omega-2i\epsilon)}\left(e(0) - \int_{-\infty}^0 dt\,e^{i(\omega-2i\epsilon)t}(e(t)e^{-2\epsilon t})'\right). \tag{47}$$

Note that $|\omega-2i\epsilon|^{-1} \leq \frac{1+\epsilon^{-1}}{1+|\omega|}$ when $\mathrm{Im}\,\omega \leq \epsilon$, we have

$$|\tilde{e}(\omega)| \leq \frac{C_2}{1+|\omega|}, \quad \mathrm{Im}(\omega) \leq \epsilon, \tag{48}$$

where

$$C_2 := (1+\epsilon^{-1})\left(|e(0)| + \epsilon^{-1}\sup_{t<0}\left|(e(t)e^{-2\epsilon s})'\right|\right) < \infty, \tag{49}$$

where $\sup_{t>0}\left|(e(t)e^{-2\epsilon t})'\right| < \infty$ because we have assumed that $|e(t)e^{-2\epsilon t}|$ and $|e'(t)e^{-2\epsilon t}|$ are bounded function.

We're now ready to derive the bound for $f(\omega)$. From (37) and (45), we know that $|f| < C'(1+|\omega|)^2|F_+(\omega)|$ for $\mathrm{Im}(\omega) \geq \epsilon$. (43) further tells us that $|F_+(\omega)| \leq \frac{e^C}{1+|\omega|}$ for $\mathrm{Im}(\omega) \geq \epsilon$, which leads to the conclusion that $|f| \leq C_3(1+|\omega|)$ for $\mathrm{Im}(\omega) \geq \epsilon$ and some constant $C_3 > 0$. Similarly, for $\mathrm{Im}(\omega) \leq \epsilon$, it's not hard to prove that $|f| \leq C_4(1+|\omega|)$ using (38) and (48). Therefore, $f(\omega)/(1+|\omega|)$ is bounded on the entire complex plane. Since $f$ is holomorphic, it has to be linear in $\omega$. Setting $f = \kappa - i\omega$ allows us to solve equation (39), and the solution is $\kappa = \frac{\alpha-H}{\alpha+H}$, with $\alpha$ given by

$$\alpha = 2\lambda(1+s_0^2)|F_+(i)|^2 > 0. \tag{50}$$

The solution for $\tilde{\xi}(\omega)$ is

$$\tilde{\xi}(\omega) = \frac{(\kappa-i\omega)}{(s_0^2-\omega_+^2)(1-i\omega)F_+(\omega)}. \tag{51}$$

Similar to the case of a rational kernel, the phase shift $\varphi$ is determined by the residue at $\omega = -s_0$. Employing the asymptotic expansion $\xi(t) \sim A e^{i s_0 t} + c.c.$, we have $A = (s_0 - i\kappa)\beta$, and the constant $\beta$ is given by

$$\beta = \frac{1}{2 s_0 (1 + i s_0) F_+(-s_0)}, \tag{52}$$

which completes the discussion for generic kernel $G(t)$. The relation between $H$ and $\varphi$ is given by

$$H = \alpha \left( \frac{1 + s_0 \tan(\varphi - \varphi_0)}{1 - s_0 \tan(\varphi - \varphi_0)} \right), \tag{53}$$

where the constant $\varphi_0$ is given by

$$\varphi_0 = \arg \left( \frac{F_+(s_0)}{1 + i s_0} \right). \tag{54}$$

Note that (50) and (54) only requires information about $F_+(i)$ and $F_+(s_0)$ to compute the phase shift $\varphi$. From (40) we have

$$\alpha = 2\lambda(1 + s_0^2) \exp \left( \frac{1}{\pi} \int_{\mathbb{R}} d\omega \, \frac{\ln F(\omega)}{1 + \omega^2} \right), \tag{55}$$

Similarly, for $\varphi_0$ we have

$$\varphi_0 = \arg \left[ \frac{1}{1 + i s_0} \exp \left( \frac{s_0}{\pi i} \, \mathrm{P.} \int_0^\infty d\omega \, \frac{\ln F(\omega)}{\omega^2 - s_0^2} \right) \right] \mod \pi, \tag{56}$$

where P. denotes the principal value. Note that the integral expression (55) and (56) only depends on the data of $F(\omega)$ on the real line, and are therefore easier to compute when $\mathcal{G}(\omega)$ is only known numerically.

## III. NON-ZERO BOUND STATE ENERGY

We analyze the case of non-zero bound state energy $E$ by solving Eq. (9). Defining $\kappa_* := \sqrt{-E}$, we consider $\kappa_* \ll \Lambda$. The substitution $\phi(k) = k^{-1}\xi(k)$ with $t \equiv \ln(\Lambda/k)$ becomes inconvenient for $E \neq 0$ due to the kernel $\sqrt{q^2 - 4E/3}$. Instead, we introduce:

$$k = \Lambda a \sinh \tau, \quad a \equiv 2\sqrt{-E/3\Lambda^2}, \quad \tau \in [0, \mathrm{arcsinh}(a^{-1})], \tag{57}$$

which gives the integral equation:

$$\lambda \xi(\tau) = \int_0^T d\sigma \left( K(\tau, \sigma) + a^2 H \sinh \tau \sinh \sigma \right) \xi(\sigma), \tag{58}$$

where $a \sinh T = 1$, $\lambda = \sqrt{3}\pi/8$, and:

$$K(\tau, \sigma) = \frac{1}{2} \ln \left( \frac{\sinh^2 \tau + \sinh^2 \sigma + \sinh \tau \sinh \sigma + 3/4}{\sinh^2 \tau + \sinh^2 \sigma - \sinh \tau \sinh \sigma + 3/4} \right). \tag{59}$$

### A. Asymptotic Analysis

For $e^T \gg 1$, we solve (58) in two regions:

**Region I** ($e^\tau \gg 1$): Since $e^T, e^\tau \gg 1$, we have the following approximation:

$$a^2 H \sinh \tau \sinh \sigma \approx H e^{-(2T - \tau - \sigma)}. \tag{60}$$

In addition, for $e^\tau \gg 1$, the kernel $K(\tau, \sigma)$ satisfies

$$K(\tau, \sigma) \approx \frac{1}{2} \ln \left( \frac{\cosh(\tau - \sigma) + 1/2}{\cosh(\tau - \sigma) - 1/2} \right), \tag{61}$$

that is, in region I, equation (58) is equivalent to (11) under a change of coordinate $t \to T - \tau$. Based on our previous analysis of (11), we know that the solution $\xi(\tau)$ exhibits a plane-wave behavior

$$\xi(\tau) \sim A \cos(s_0(\tau - T) - \varphi), \quad 1 \ll e^\tau \ll e^T. \tag{62}$$

**Region II** ($e^{T-\tau} \gg 1$): The $H$ term is suppressed, and the solution satisfies:

$$\lambda \xi(\tau) = \int_0^\infty d\sigma \, K(\tau, \sigma) \xi(\sigma), \tag{63}$$

with asymptotic form:

$$\xi(\tau) \sim A' \cos(s_0 \tau - \eta), \quad 1 \ll e^\tau \ll e^T. \tag{64}$$

### B. Matching and the Energy Spectrum

Matching phases in the overlap region requires:

$$\eta = s_0 T + \varphi \mod \pi. \tag{65}$$

With $\varphi = s_0 \ln(\Lambda_*/\Lambda) + k\pi$ and $e^T \to 2/a$, we obtain the quantization condition:

$$s_0 \ln\left(\frac{\kappa_*}{c\Lambda_*}\right) = 0 \mod \pi, \tag{66}$$

where $c = \sqrt{3} e^{-\eta/s_0}$. This yields an infinite tower of bound states:

$$E_n = -(c\Lambda_*)^2 e^{-2\pi n/s_0}, \quad -E_n/\Lambda^2 \ll 1. \tag{67}$$

#### 1. Exact Solution

We shall now show that (63) is exactly solvable with $\eta = \pi/2 \mod \pi$. To see this, it's central to notice the following factorization:

$$\frac{\sinh^2(\tau) + \sinh^2(\sigma) + \sinh(\tau)\sinh(\sigma) + 3/4}{\sinh^2(\tau) + \sinh^2(\sigma) - \sinh(\tau)\sinh(\sigma) + 3/4} = \left(\frac{\cosh(\tau - \sigma) + 1/2}{\cosh(\tau - \sigma) - 1/2}\right)\left(\frac{\cosh(\tau + \sigma) - 1/2}{\cosh(\tau + \sigma) + 1/2}\right), \tag{68}$$

The factorization (68) leads to the decomposition

$$K(\tau, \sigma) = G(\tau - \sigma) - G(\tau + \sigma), \tag{69}$$

where $G(t)$ was the kernel defined in (12). It implies that eigenfunctions of $K(\tau, \sigma)$ are sine functions:

$$\int_0^\infty d\sigma \, K(\tau, \sigma) \sin(\omega\sigma) = \mathcal{G}(\omega) \sin(\omega\tau). \tag{70}$$

Thus, (63) is solved by $\xi(\tau) = A' \sin(s_0 \tau)$. Therefore, we might take $\eta = \pi/2$ and

$$c = e^{-\pi/2s_0} \sqrt{3} \approx 0.363581, \tag{71}$$

which completes our discussion.


\fi

\end{document}